\documentclass[12pt]{article}
\usepackage{graphicx}
\usepackage{amssymb}
\usepackage{scicite}
\usepackage{setspace}
\usepackage{url}
\title{Foraging under conditions of short-term exploitative competition: The case of stock traders}

\author{Serguei Saavedra$^{1}$, R. Dean Malmgren$^{2}$,\\Nicholas Switanek$^{3,4}$, Brian Uzzi$^{3,4}$\\
\scriptsize{$^1$Integrative Ecology Group, Estaci\'on Biol\'ogica de Do\~nana (EBD-CSIC)}\\
\scriptsize{Am\'erico Vespucio s/n, E-41092 Sevilla, Spain}\\
\scriptsize{$^2$Datascope Analytics, Chicago, Illinois, 60603, USA}\\
\scriptsize{$^3$Northwestern Institute on Complex Systems, Northwestern University, Evanston, Illinois, 60208, USA}\\
\scriptsize{$^4$Kellogg School of Management, Northwestern University, Evanston, Illinois, 60208, USA}}

\date{}
 
\begin{document}
%\linenumbers

%\baselineskip12pt

\maketitle 
\begin{flushleft}
\end{flushleft}
 
\begin{abstract}
Theory purports that animal foraging choices evolve to maximize
returns, such as net energy intake. Empirical research in
both human and nonhuman animals reveals that individuals often attend
to the foraging choices of their competitors while making their own
foraging choices. Due to the complications of gathering field data or
constructing experiments, however, broad facts relating theoretically
optimal and empirically realized foraging choices are only now
emerging. Here, we analyze foraging choices of a cohort of
professional day traders who must choose between trading the same
stock multiple times in a row---patch exploitation---or switching to a
different stock---patch exploration---with potentially higher
returns. We measure the difference between a trader's resource intake
and the competitors' expected intake within a short period of time---a
difference we call short-term comparative returns. We find that traders' choices can be explained by foraging heuristics that maximize their daily short-term comparative returns. However, we find no one-best relationship between different trading choices and net income intake. This suggests that traders' choices can be short-term win oriented and, paradoxically, maybe maladaptive for absolute market returns.
\end{abstract}

%\clearpage

\section{Introduction}

Animals face a recurring alternative between continuing to forage in a
patch or gambling on switching to a different patch with possibly
better returns \cite{Green,Cooper1,Cooper2}. Optimal foraging theory
purports that animal foraging choices have been shaped by natural
selection and should maximize absolute fitness
\cite{MacArthur,Emlen,Stephens,Pyke}. Similarly, optimal foraging theory considers that both human and nonhuman
animals can take into account the foraging choices of their competitors
while making their own choices
\cite{McDiarmid,Bateson,Stephens2,Houston,Pompilio,Freidin,Simon,Kahneman,Novemsky,Thaler,Haigh}. Thus, interactions
among competitors are increasingly important to understanding how real
foraging choices can be shaped as animals compete for
resources
\cite{Park,Park2,Grover,Vahl,Dall,Le,Whitehouse,Mitchell}. Competitive
interactions are typically of two types: exploitative competition, when
different animals consume common limited resources (e.g., two
predators hunting the same prey) \cite{Park,Park2,Grover}; and interference competition, when direct interactions such as territoriality negatively affect the foraging of other animals \cite{Park2,Grover,Vahl}. Yet, broad empirical facts on the link between optimal and real foraging choices are scarce due to the complication of gathering field data or constructing experiments \cite{Holt}.

Importantly, biological and socio-economic systems share many common features in terms
of distributed resources and competition, and thus financial systems have
provided a fruitful and intriguing setting to test biological theories
of behavior because of their high quality quantifiable and dynamic
behavioral data
\cite{Park2,Saavedra,Kirman,Haldane,Saavedra3,Saavedra2}. As far as we
know, however, financial traders have not been examined from the perspective of foraging. Day traders face the classical foraging trade-off of trading the same stock multiple times in a row---patch exploitation---or switching to a different stock---patch exploration. For instance, each trader can trade multiple stocks within a class of stocks she has expertise in (e.g., technology stocks, banks stocks, transportation stocks, etc.) and is faced with the foraging choice of buying and selling the same stock multiple times in a row (e.g. buy a stock at a low price and vice versa for selling) or switching their trading to a different stock where returns are potentially higher (Fig. 1). By analogy to foraging in a physical habitat where energy is invested in traveling and hunting, traders either exploit the returns related to one stock (i.e., a patch) or explore a different patch while potentially experiencing cognitive costs for switching between patches \cite{Hamilton,Danzinger,Monsell,Minear}. Moreover, the returns in each patch are shaped by exploitative competition, where the foraging choices of other traders, even within a short period of time, can increase or decrease the quality and availability of resources as they choose to buy or sell their stocks \cite{Moro}. Thus, if a trader is willing to buy and the majority of traders are also buying then the stock price increases, in turn, the trader's return will be reduced. In this paper we investigated the extent to which professional traders' exploration and exploitation choices can be explained by foraging heuristics that respond to short-term competition with other traders. Additionally, we analyzed whether traders' trading choices are associated with their net income intake. A significant relationship would mean a real correspondence between trading choices and absolute returns; whereas a lack of relationship would suggest a maladaptive behavior for absolute market returns.

\section{Material and methods}

We studied the second-by-second trading decisions of day traders at a typical small-to-medium sized trading firm from January 1, 2007 to December 31, 2008. We recorded when a trader begins to trade a stock, how much he subsequently traded the same stock, and when he switched to explore a different stock. In our data, traders typically ($>90\%$ of the time) made more than $10$ transactions, and more than $3$ switches, per day (Fig. 2). These novel data cover more than 300 thousand trades made on approximately $3000$ different stocks across a very wide range of sectors and on various exchanges, mostly from NYSE, the ``blue chip" exchange, and NASDAQ, the exchange known for high tech and volatile stocks. In particular, the stocks include high technology firms, diversified financials, shipping, natural resources, construction, chemicals, insurance, steel, etc. The top 5 stocks traded at the firm over our time period in terms of number of trades and volume were JP Morgan Chase \& Co., Mechel Steel Group Oao, Goldman Sachs Group, Apple Inc., and Potash Corporation of Saskatchewan Inc.

A typical small-to-medium day trading firm invests the money of the owners of the firm in stocks and hires traders to make the firm's investments. Day traders make only intraday trades; they typically do not hold inventories of stocks beyond a single day. Rather, they enter and exit positions each day during normal trading hours of 9:30 AM and 4:00 PM (EST). Our day traders are ``point-and-clickers." They make trades in real time 98\% of the time (the 1.2\% of the trades done algorithmically were omitted and did not affect the results). Though they sit in the same firm, day traders typically trade different stocks from each other and trade independently of each other. Trading different stocks diversifies the firm's holdings, exploits specialized trading knowledge, and avoids accidentally trading against each other's positions. These dynamics mean that traders have little incentive to mimic each other's trades, information gathering behavior, or trading decisions. The firm was located in the US. 

Our sample of day traders under study was 30. This sample of 30 traders was the full number of traders for which there was complete data on all decisions and behaviors measured over our observation period. By contrast, the other traders at the firm ($n=36$) all worked for truncated interludes or worked erratically, which made their measurement unbalanced and unsystematic, and vulnerable to selection and small sample size biases \cite{Gardner}. All traders at the firm were men of an average age of 35 years old and a range between 22 and 50 years of age. They used the same technology to trade, had access to the same public information sources, and were subject to an equivalent incentive scheme.  Traders were paid a base salary plus commissions on trades. The firm did not share with us their commission formula. They did indicate that like typical firms, the commission was based on end of the day earnings over a range of time to remove as possible chance fluctuations.

At the time of observation, our sample of traders traded about half of the stocks available on these exchanges on average. It is likely that the specific company stocks that were not traded were ones that lent themselves to holding long-term positions rather than trading on intraday shifts in price. All trading related data was automatically captured by the firm's trading system, which is specially designed for accuracy in recording, and used by most other firms in the industry. This automated and electronic capture system works unobtrusively to avoid interference with trading. The capture system fulfills US Securities and Exchange Commission requirements that all trades be recorded and archived for up to 7 years. The net income data were calculated by the firm using standard industry metrics. In our study, we analyzed all the trades of all the stocks of all the traders in our sample. The study conforms to Institutional Review Board (IRB) criteria.  There was no subject interaction, all data was 100\% archival, and the firm and the subjects were anonymized. Legally, all data used in the study is owned by the company. All traders at the firm know the firm owns the data and that their communications and trading behavior is recorded by law. We received written permission from the firm to use these data for research purposes and publishing contingent on identifying characteristics of the firm and its traders remaining confidential and anonymous.

\section{Results}

\subsection*{{\it (a) Short-term comparative returns}}

To measure the extent to which traders' exploration and exploitation choices can be shaped by the foraging choices of their competitors, we introduced a novel measure that captures the difference between a trader's resource intake and competitors' expected intake over a short period of time---what we called short-term comparative return---and tested whether foraging choices can be explained by traders trying to maximize their daily short-term comparative returns. 

The short-term comparative return associated with each transaction was
calculated as the difference between actual traded prices decided on
by each trader and the average prices in the market within a relevant
time window. Since the anticipation of and response to the actions of
competitors can be manifested by acting before them or by waiting and
acting after them \cite{Carmerer,Sutton,Kuhnen,Zhu}, we followed
theory and defined context limits according to the smallest time
window (5 minutes) where it has been shown that individual
transactions can impact the returns of others in the market
\cite{Moro}. For each trader $i$ and each of his transactions $j$ on
day $t$, we defined the short-term comparative return as
$z_{ijt}=(\gamma)(T_{ijt}- \langle
T_{st}^*\rangle)/\sigma_{T_{st}^*}$, where $T_{ijt}$ is the traded
price and $\langle T_{st}^*\rangle$ and $\sigma_{T_{st}^*}$ are the
average and standard deviation of stock $s$'s price on day $t$ within
a five-minute interval, and $\gamma=1$ ($\gamma=-1$) for selling
(buying) transactions (Fig. 1). The stock's average price $\langle
T_{st}^*\rangle$ and standard deviation $\sigma_{T_{st}^*}$ are a
mirror of the foraging choices of competitors, since prices move
according to the stock's consumption or demand
\cite{Moro,Kirman}. These price statistics are computed using the WRDS database \cite{WRDS}, which has all the recorded transactions made around the world for each stock. Thus $z_{ijt}>0$ and $z_{ijt}<0$ always indicate, respectively, a positive and negative short-term comparative return relative to the actions of competitors at that time. 

To test how well the time window captures the changing foraging
choices and depletion of resources over a period of time, we
calculated lagged and leading short-term comparative returns
$z_{ijt}(\Delta)$ using the stock's average price $\langle
T_{st}^*(\Delta)\rangle$ and standard deviation
$\sigma_{T_{st}^*}(\Delta)$ within five-minute intervals 5 minutes
before $\Delta^-$ and respectively 5 minutes after $\Delta^+$ the observed five-minute interval of each transaction. Again, we used the WRDS database \cite{WRDS} to calculate these values. If the distribution of lagged short-term comparative returns is similar to the actual distribution of short-term comparative returns then it would suggest that the prices within the actual time window are, in fact, representative of the actions of others over a recent short period of time and not simply artifacts of the five-minute interval. Using Wilcoxon signed rank test for testing paired and non-normally distributed distributions, we found that in 28 out of 30 traders the actual and lagged short-term comparative returns were significantly similar (Table 1), which confirms that the actual time window is a reasonable context to use. Additionally, we repeated the same analysis but with lags ($\Delta^-$,$\Delta^+$) greater than 1 hour and found in all traders the actual and lagged returns were significantly different (Table 1), meaning that these prices are representative of the actions of others only in the short-term.

To know whether traders' short-term comparative returns are associated with their foraging choices, we divided the total number $N_{it}$ of transactions $j$ of each trader $i$ in day $t$ according to their exploration index $b_{ijt}$, or trading patch, and their exploitation index $q_{ijt}$, or position within the patch. Figure 1 presents an illustrative example of how we divided the number of transactions. This example shows that a trader $i$ in a day $t$ had a total of 14 transactions (green bars) allocated in 4 different patches (gray regions). Regarding the exploration index, the first two transactions were characterized by $b_{ijt}=1$ for $j=\{1,2\}$, the next 4 transactions by $b_{ijt}=2$ for $j=\{3,4,5,6\}$, the next 6 transactions by $b_{ijt}=3$ for $j=\{7,8,9,10,11,12\}$, and the last 2 transactions by $b_{ijt}=4$ for $j=\{13,14\}$. Additionally, these transactions were characterized by their exploitation indices $q_{i1t}=1$, $q_{i2t}=2$, $q_{i3t}=1$, $q_{i4t}=2$, $q_{i5t}=3$, $q_{i6t}=4$, $q_{i7t}=1$, $q_{i8t}=2$, $q_{i9t}=3$, $q_{i10t}=4$, $q_{i11t}=5$, $q_{i12t}=6$, $q_{i13t}=1$ and $q_{i14t}=2$. Note that each time the trader visits a new patch, the exploitation index is reset to 1.

For each trader, we modeled short-term comparative returns as a function of the importance of exploration $\beta_1$ and exploitation $\beta_2$ using a multivariate regression model that takes the form $z_{ijt}=\beta_0+\beta_1b_{ijt}+\beta_2q_{ijt}+\epsilon$. Table 1 indicates that both exploration and exploitation are negatively associated with short-term comparative returns. In line with optimal foraging theory, these results reveal diminishing payoffs per resource, i.e., daily comparative returns $R_{it}=\sum_{j=1}^{j=N_{it}}z_{ijt}$ can decrease in proportion of the number of stocks exploited or explored.

To illustrate this point, we used $\beta_1=-0.002$ and $\beta_2=-0.02$ of one single trader, and assuming that in one particular day that trader made $65$ transactions exploring 65 different patches, the trader would have $\hat{R}_{it}=\sum_{j=1}^{j=65}\hat{z}_{ijt}=-5.59$, where $\hat{z}_{ijt}$ are the predicted short-term comparative returns from the regression model without considering $\beta_0$, i.e., this decline is relative to the trader's average returns over the same transactions. In contrast, if the trader would have explored one single patch, the total returns would have changed to $\hat{R}_{it}=-43.03$. If one multiplies $\hat{R}_{it}$ by say the average difference between traded price and average price in the market ($\$0.13$) times the average volume of stocks per transaction ($300$) in our data, $\hat{R}_{it}$ translates into a {\emph relative} loss compared to the trader's average performance over the same transactions of $\$-218.01$ and $\$-$1678.17, respectively. Note that this negative return is a relative measure of performance and should not be interpreted as the actual payoff. Instead, it reflects the possibility of different expected outcomes \cite{Rode}.  

This resulting relative loss indicates that when foraging is
compounded over many choices of exploitation and exploration,
different activity patterns can impact the daily short-term
comparative returns of traders. Similarly, the relative loss can also be examined
by quantifying the decline in $\hat{R}_{it}$ generated by exploitation
and exploration patterns separately when considering a constant number
of transactions per patch $Q_{it}$. Figure 3 shows that when
considering exploration only, the lower the value of $Q_{it}$ the
higher the decline of $\hat{R}_{it}$ (dashed line); and the opposite
behavior is observed when considering exploitation only (solid
line). Importantly, the relationship between exploitation and
exploration patterns reveals that an optimal pattern for jointly maximizing traders' $\hat{R}_{it}$ exists, i.e. the intersection between the two curves.

\subsection*{{\it (b) Optimal short-term comparative returns}}

To test whether traders' foraging choices respond to maximize their daily short-term comparative returns, we measured the extent to which the observed number of transactions per patch $Q_{it}$ agreed with the optimal transactions per patch $Q^*_{it}$. To find $Q^*_{it}$, we used the equality of returns from the exploration and exploitation curves to describe the intersection point of the curves in order to then estimate the expected optimal number of transactions. Mathematically, we calculated the value that maximizes $\hat{R}_{it}$ given by $(\frac{Q^*_{it}}{2})(Q^*_{it}+1)(\frac{\langle N_{it}\rangle \beta_2}{Q^*_{it}})=(\frac{\langle N_{it}\rangle}{2Q^*_{it}})(\frac{\langle N_{it}\rangle}{Q^*_{it}}+1)(Q^*_{it}\beta_1)$, where $\langle N_{it}\rangle$ is the mean number of total transaction of trader $i$, and $\beta_1$ and $\beta_2$ are, respectively, the importance of exploration and exploitation taken from the multivariate regression model for each trader separately (Table 1). Thus, the expected optimal number of transaction per patch is the positive root of $(Q^*_{it})^2+Q^*_{it}(1-\frac{\beta_1}{\beta_2})-\frac{\langle N_{it}\rangle \beta_1}{\beta_2}$.

Interestingly, we found that exploration and exploitation choices can, in fact, be explained by traders trying to maximize their daily short-term comparative returns. We measured the deviation between the optimal $Q^*_{ij}$ and the distribution of actual values of $Q_{ij}$ using the normalized model error (NME) for each individual case \cite{Pires}. Here, the NME was computed as the difference between $Q^*_{ij}$ and the observed median value of $Q_{ij}$ divided by the difference between the observed median value and the observed value of $Q_{ij}$ at the $2.5\%$ or $97.5\%$ quantiles, depending on whether the optimal value is lower or larger than the observed median value. The NME makes no particular assumption about the distribution of observed values. NME values between $(-1,1)$ can be taken as cases where the optimal value is significantly similar to the observed values \cite{Pires}. We found only 3 cases with NME values greater than 1 (Fig. 4). Importantly, this number of cases falls within the number of rejections $[0,4]$ that one would expect with $95\%$ confidence from a Binomial model $B(30,0.05)$ \cite{Dean}. Thus, one cannot reject the hypothesis that this model is a good approximation to the observed exploration and exploitation choices of traders. Broadly, our findings reveal that traders' choices can be explained by foraging heuristics that maximize their daily short-term comparative returns.

Finally, to test whether traders' choices are associated with their net income intake, we introduced two additional return metrics. The first metric, which we called actual relative return $A_{it}$, provides information about the amount of money made by traders relative to the expected amount made by competitors. It is calculated similar to the short-term comparative returns measure except that it does not take into account the standard deviation; and instead, it multiplies returns by the number of stocks sold or bought.  The second metric, which we called net income intake, $I_{it}$, is simply the amount made by traders; it is does not compare it with competitors. Figure 5A shows a significant and positive association between short-term comparative returns and actual relative returns, confirming that traders' choices respond to short-term competition with other traders. In contrast, Figure 5B shows no association between comparative returns and net income intake, revealing a significant deviation between traders' short-term returns and their absolute returns. This suggest that traders' potential focus on short-term competition may come at the cost of missing net income optimizing opportunities.

\section{Discussion} 

Optimal foraging theory has proven useful for understanding how the
fitness and survivability of animals depends on the trade-off between
effort expended and absolute resources gained. It has further been
shown that human and nonhuman animals rarely make the core foraging
trade-off independently: their foraging choices are influenced by the
choices their competitors make. Nonetheless, the study of the
relationship between optimal and real foraging choices remains
nascent. Here we investigated whether the exploration and
exploitation choices of day traders can be explained by short-term
exploitative competition. Traders' foraging choices may be more
abstract, stochastic, and rapid than foraging choices in physical
environments, yet the same mechanisms may underpin the allocation of vast financial and material resources under competition \cite{Saavedra,Coates,Hill,Payne,McDermott}. 

Our study analyzed the investing choices made by a
cohort of 30 day traders at one firm. By analogy to foraging in the
physical world, these traders sought to find the most beneficial
compromise between the costs and benefits of continued foraging within
a patch (i.e., consecutively buying and selling of the same stock) or
switching to forage in a new patch (i.e., trading a different stock),
where the returns to trading are affected by the foraging choices made
by competitors. We measured traders' short-term comparative returns as
the marginal difference between their actual returns to trading a
stock and the mean returns possible based on the competitors' foraging
choices in the market within a relevant period of time. We found that
traders' short-term comparative returns are subject to an important
trade-off between exploration and exploitation. We could not reject
the hypothesis that traders' exploration and exploitation choices can
be explained by traders following short-term choices that focus on maximizing their daily short-term
comparative returns. While a complete determination of the drivers of
these choices is beyond our analysis, one possible account for the
observed behavior is that traders first visit the patch in which they
do best, then next best, and so on. Thus, traders may choose patches
that descend in worth, assuring at least early success, while limiting
exposure to unpredictable shifts in competition in a patch that might
create losses for the trader \cite{Rode}. Such trading choices, however, may be different under new algorithmic trading where price transactions are previously fixed \cite{Moro}. 

Foraging animals appear to optimally decide what patch of resources will offer the best returns to their efforts and how long to stay in a patch before moving onto the next best patch. Remarkably, our findings revealed that stock traders' trading choices can be explained by similar foraging heuristics that respond to short-term competition with other traders. However, there were important differences too. We found no one-best relationship between different trading choices and net income intake, suggesting that traders' choices can be short-term win oriented and, paradoxically, maybe maladaptive for absolute market returns \cite{Barber1,Barber2}. This implies that traders' net income intake might be more strongly associated to global outcomes, social contagion, or sporadic big losses and wins \cite{Kirman,Saavedra,Saavedra3}. While the same problem is not true of animal foraging since the resources gained from each patch are also the net payoffs, it would be interesting to investigate whether maladaptive foraging behavior can arise under rapid changing environments. In financial settings, it remains to see the extent to which this deviation between short-term choices and net income intake can influence the instability of markets.

\medskip
\singlespacing
\section*{Acknowledgements} 
We thank Alex Bentley, Esteban Freidin, Cristi\'{a}n Huepe, Rudolf Rohr, and Michael Schnabel for useful comments on a previous draft. Funding was provided by the Kellogg School of Management, Northwestern University, the Northwestern University Institute on Complex Systems (NICO), and the Army Research Laboratory under Cooperative Agreement W911NF-09-2-0053. NSF VOSS grant (OCI-0838564). SS also thanks CONACYT.

\bibliographystyle{Science}

\clearpage

\addtolength{\hoffset}{-2cm}

\begin{table}
	\centering
	\scriptsize
		\begin{tabular}{llllllllll}
			Trader&$n$&$\langle N_{it}\rangle$&$z_{5min}$&$z_{1hr}$&$B_0$&$B_1$&$B_2$&$\rho_{rel}$&$\rho_{abs}$ \\ \hline1	&	3420	&	32.3	&	$	0.45	$	&	$	-10.724	$	&	$	-0.144	(	-0.0681	)	^	{	**	}	$	&	$	-0.002	(	0.0006	)	^	{	*	}	$	&	$	-0.021	(	0.0029	)	^	{	*	}	$	&	$	0.74	^	{	***	}	$	&	$	0.141	^	{		}	$	\\
2	&	5049	&	59.9	&	$	1.16	$	&	$	-9.659	$	&	$	-0.095	(	0.0069	)	^	{	*	}	$	&	$	-0.002	(	0.0005	)	^	{	**	}	$	&	$	-0.017	(	0.0016	)	^	{	*	}	$	&	$	0.711	^	{	***	}	$	&	$	-0.029	^	{		}	$	\\
3	&	10112	&	37.0	&	$	-0.46	$	&	$	6.032	$	&	$	-0.158	(	0.0231	)	^	{	***	}	$	&	$	-0.001	(	0.0013	)	^	{		}	$	&	$	-0.002	(	0.0002	)	^	{	*	}	$	&	$	0.527	^	{	***	}	$	&	$	0.062	^	{		}	$	\\
4	&	3324	&	37.3	&	$	-0.42	$	&	$	-7.001	$	&	$	-0.085	(	0.0424	)	^	{	**	}	$	&	$	-0.002	(	0.0002	)	^	{	*	}	$	&	$	-0.023	(	0.0045	)	^	{	**	}	$	&	$	0.597	^	{	***	}	$	&	$	-0.101	^	{		}	$	\\
5	&	27278	&	87.0	&	$	-1.99	$	&	$	10.23	$	&	$	0.05	(	0.0206	)	^	{	**	}	$	&	$	-0.003	(	0.0004	)	^	{	***	}	$	&	$	-0.011	(	0.0022	)	^	{	***	}	$	&	$	0.44	^	{	***	}	$	&	$	-0.007	^	{		}	$	\\
6	&	36869	&	105.2	&	$	0.76	$	&	$	8.007	$	&	$	0.254	(	0.0305	)	^	{	***	}	$	&	$	-0.001	(	0.0004	)	^	{	***	}	$	&	$	-0.008	(	0.0031	)	^	{	***	}	$	&	$	0.59	^	{	***	}	$	&	$	0.1	^	{		}	$	\\
7	&	32345	&	107.1	&	$	0.99	$	&	$	14.146	$	&	$	-0.28	(	0.0061	)	^	{	***	}	$	&	$	-0.001	(	0.0001	)	^	{	***	}	$	&	$	-0.009	(	0.0004	)	^	{	**	}	$	&	$	0.505	^	{	***	}	$	&	$	0.042	^	{		}	$	\\
8	&	31290	&	95.2	&	$	0.83	$	&	$	-6.345	$	&	$	0.018	(	0.0101	)	^	{	*	}	$	&	$	-0.005	(	0.0001	)	^	{	**	}	$	&	$	-0.002	(	0.0129	)	^	{		}	$	&	$	0.322	^	{	***	}	$	&	$	0.279	^	{	***	}	$	\\
9	&	7359	&	81.3	&	$	0.23	$	&	$	-12.343	$	&	$	-0.301	(	0.0571	)	^	{	***	}	$	&	$	-0.001	(	0.0002	)	^	{	*	}	$	&	$	-0.012	(	0.0019	)	^	{	*	}	$	&	$	0.551	^	{	***	}	$	&	$	-0.055	^	{		}	$	\\
10	&	5671	&	46.8	&	$	-1.35	$	&	$	13.234	$	&	$	0.208	(	0.0379	)	^	{	***	}	$	&	$	-0.002	(	0.0005	)	^	{	***	}	$	&	$	-0.25	(	0.0081	)	^	{	***	}	$	&	$	0.161	^	{	**	}	$	&	$	-0.008	^	{		}	$	\\
11	&	3752	&	36.2	&	$	0.38	$	&	$	-9.811	$	&	$	0.482	(	0.0316	)	^	{	***	}	$	&	$	-0.003	(	0.0008	)	^	{	*	}	$	&	$	-0.004	(	0.0011	)	^	{	*	}	$	&	$	0.567	^	{	***	}	$	&	$	-0.042	^	{		}	$	\\
12	&	7478	&	74.2	&	$	0.68	$	&	$	5.046	$	&	$	-0.281	(	0.0281	)	^	{	***	}	$	&	$	-0.001	(	0.0004	)	^	{	***	}	$	&	$	-0.004	(	0.0012	)	^	{	**	}	$	&	$	0.524	^	{	***	}	$	&	$	0.091	^	{		}	$	\\
13	&	18835	&	60.9	&	$	0.12	$	&	$	-14.345	$	&	$	-0.171	(	0.016	)	^	{	***	}	$	&	$	-0.003	(	0.0009	)	^	{	***	}	$	&	$	-0.022	(	0.001	)	^	{	*	}	$	&	$	0.418	^	{	***	}	$	&	$	-0.064	^	{		}	$	\\
14	&	9690	&	64.2	&	$	-2.21	$	&	$	-8.549	$	&	$	-0.21	(	0.0214	)	^	{	***	}	$	&	$	-0.005	(	0.0005	)	^	{	**	}	$	&	$	-0.003	(	0.0018	)	^	{	*	}	$	&	$	0.414	^	{	***	}	$	&	$	0.044	^	{		}	$	\\
15	&	8845	&	56.1	&	$	-0.44	$	&	$	-3.408	$	&	$	-0.115	(	0.0205	)	^	{	***	}	$	&	$	-0.002	(	0.0006	)	^	{	***	}	$	&	$	-0.018	(	0.0026	)	^	{	***	}	$	&	$	0.49	^	{	***	}	$	&	$	-0.023	^	{		}	$	\\
16	&	22289	&	95.5	&	$	-1.79	$	&	$	-8.786	$	&	$	0.49	(	0.034	)	^	{	***	}	$	&	$	-0.013	(	0.0032	)	^	{	***	}	$	&	$	-0.098	(	0.0004	)	^	{	***	}	$	&	$	0.318	^	{	***	}	$	&	$	-0.371	^	{	***	}	$	\\
17	&	19712	&	66.1	&	$	-0.79	$	&	$	-9.125	$	&	$	0.038	(	0.0026	)	^	{	*	}	$	&	$	-0.004	(	0.0019	)	^	{	**	}	$	&	$	-0.007	(	0.0014	)	^	{	***	}	$	&	$	0.669	^	{	***	}	$	&	$	-0.288	^	{	***	}	$	\\
18	&	4938	&	47.8	&	$	1.11	$	&	$	13.419	$	&	$	-0.222	(	0.0282	)	^	{	***	}	$	&	$	-0.002	(	0.0008	)	^	{	***	}	$	&	$	-0.019	(	0.004	)	^	{	***	}	$	&	$	0.748	^	{	***	}	$	&	$	0.002	^	{		}	$	\\
19	&	20661	&	66.3	&	$	-1.87	$	&	$	14.218	$	&	$	0.006	(	0.021	)	^	{		}	$	&	$	-0.002	(	0.0006	)	^	{	***	}	$	&	$	-0.076	(	0.0055	)	^	{	***	}	$	&	$	0.233	^	{	***	}	$	&	$	0.141	^	{		}	$	\\
20	&	2923	&	38.1	&	$	-1.52	$	&	$	-4.528	$	&	$	-0.243	(	0.1078	)	^	{	**	}	$	&	$	-0.003	(	0.0019	)	^	{	*	}	$	&	$	-0.006	(	0.0028	)	^	{	*	}	$	&	$	0.456	^	{	***	}	$	&	$	0.012	^	{		}	$	\\
21	&	11714	&	44.2	&	$	2.99	$	&	$	-9.342	$	&	$	-0.237	(	0.0177	)	^	{	**	}	$	&	$	-0.005	(	0.0008	)	^	{	*	}	$	&	$	-0.04	(	0.019	)	^	{	*	}	$	&	$	0.078	^	{		}	$	&	$	-0.087	^	{		}	$	\\
22	&	6123	&	51.3	&	$	-0.66	$	&	$	-6.695	$	&	$	-0.209	(	0.0285	)	^	{	***	}	$	&	$	-0.001	(	0.0006	)	^	{	*	}	$	&	$	-0.023	(	0.0061	)	^	{	**	}	$	&	$	0.655	^	{	***	}	$	&	$	-0.072	^	{		}	$	\\
23	&	10924	&	30.8	&	$	-0.80	$	&	$	11.989	$	&	$	0.032	(	0.0137	)	^	{	*	}	$	&	$	-0.003	(	0.0013	)	^	{	**	}	$	&	$	-0.015	(	0.0059	)	^	{	**	}	$	&	$	0.465	^	{	***	}	$	&	$	-0.544	^	{	***	}	$	\\
24	&	2432	&	9.3	&	$	1.42	$	&	$	-8.487	$	&	$	0.377	(	0.0885	)	^	{	**	}	$	&	$	-0.001	(	0.0003	)	^	{	**	}	$	&	$	-0.034	(	0.013	)	^	{	***	}	$	&	$	0.713	^	{	***	}	$	&	$	-0.101	^	{		}	$	\\
25	&	3987	&	40.2	&	$	-1.42	$	&	$	4.476	$	&	$	0.165	(	0.0421	)	^	{	***	}	$	&	$	-0.009	(	0.0022	)	^	{	**	}	$	&	$	-0.077	(	0.011	)	^	{	***	}	$	&	$	0.553	^	{	***	}	$	&	$	0.002	^	{		}	$	\\
26	&	3014	&	83.2	&	$	0.12	$	&	$	-7.729	$	&	$	0.395	(	0.0378	)	^	{	***	}	$	&	$	-0.001	(	0.0004	)	^	{	***	}	$	&	$	-0.023	(	0.0025	)	^	{	***	}	$	&	$	0.921	^	{	***	}	$	&	$	0.566	^	{		}	$	\\
27	&	5125	&	22.3	&	$	-1.63	$	&	$	-4.325	$	&	$	-0.046	(	0.0081	)	^	{	*	}	$	&	$	-0.003	(	0.0005	)	^	{	*	}	$	&	$	-0.026	(	0.0111	)	^	{	*	}	$	&	$	0.942	^	{	***	}	$	&	$	-0.311	^	{		}	$	\\
28	&	3947	&	44.6	&	$	1.38	$	&	$	9.987	$	&	$	0.001	(	0.0953	)	^	{		}	$	&	$	-0.002	(	0.0009	)	^	{	*	}	$	&	$	-0.011	(	0.0062	)	^	{	*	}	$	&	$	0.598	^	{	***	}	$	&	$	-0.056	^	{		}	$	\\
29	&	4921	&	55.8	&	$	1.45	$	&	$	8.696	$	&	$	0.152	(	0.0811	)	^	{	**	}	$	&	$	-0.002	(	0.0007	)	^	{	*	}	$	&	$	-0.012	(	0.0045	)	^	{	*	}	$	&	$	0.176	^	{	*	}	$	&	$	0.252	^	{	**	}	$	\\
30	&	2848	&	72.1	&	$	-1.34	$	&	$	3.198	$	&	$	0.633	(	0.0688	)	^	{	***	}	$	&	$	-0.015	(	0.0067	)	^	{	*	}	$	&	$	-0.045	(	0.0032	)	^	{	*	}	$	&	$	0.944	^	{	**	}	$	&	$	-0.771	^	{	***	}	$	\\ \hline
		\end{tabular}
\caption{Traders' detailed information. For each trader, the table shows the total number transactions made $n$ and the mean number of daily transactions $\langle N_{it}\rangle$ over the observation period. The Wilcoxon signed rank tests $z_{5min}$ and $z_{1hr}$ for lags of 5 mins and 1hr respectively. Note that values of $z>|2|$ are considered statistically significant. The coefficients $B_0$, $B_1$ and $B_2$ taken from the multivariate regression model that takes the form $z_{ijt}=\beta_0+\beta_1b_{ijt}+\beta_2q_{ijt}+\epsilon$. $(\cdot)$ corresponds to standard errors. The correlation values $\rho_{rel}$ and $\rho_{abs}$ correspond, respectively, to the association of daily short-term comparative returns $R_{it}$ with actual relative returns $A_{it}$ and net income intake $I_{it}$ (see text). $^{*}$, $^{**}$ and $^{***}$ correspond, respectively, to statistical significance levels of 10, 5 and 1 percent. Calculations are performed with software Stata.
}
\label{table1}
\end{table}

%\section{Figure Legends}

\clearpage	

\addtolength{\hoffset}{2cm}

\begin{figure*}
\centerline{\includegraphics*[width=5.5in]{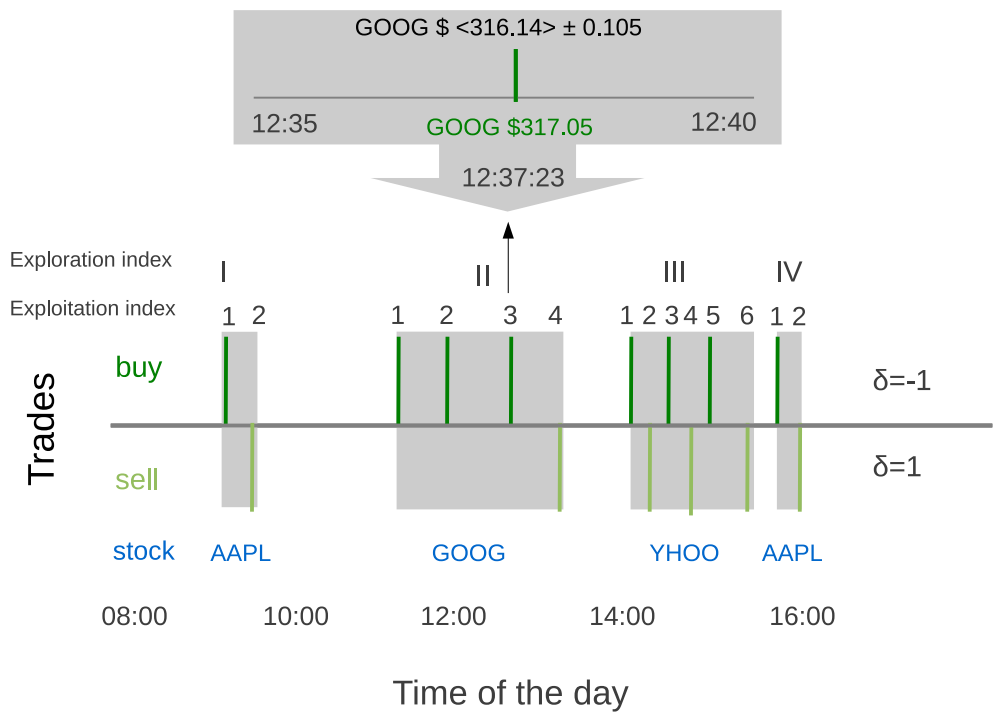}}
\caption{Foraging choices. The figure shows an illustrative example of a trading activity from a single trader. Gray boxes correspond to the different trading patches (Roman numerals) of sequential transactions of the same stock. Note that patches are separated when two consecutive transactions belong to a different stock. Arabic numerals represent the exploitation index of each transaction within its trading patch. The upper gray region is a zoom to transaction 3 within trading patch II. This transaction took place at 12:37:23 hrs, where the trader bought GOOG stock at $317.05$ USD. The stock's market price during the 5-minute window between 12:35 hrs and 12:40 hrs was $\langle 316.14\rangle \pm 0.105$(s.d.). Therefore, the short-term comparative return (see text) of this transaction can be calculated as $z_{ijt}=(-1)\frac{317.05-316.14}{0.105}=-10.38$. This suggests that this was a negative short-term comparative return for that time and choice in specific.}
\label{fig1}
\end{figure*}

\begin{figure*}
\centerline{\includegraphics*[width=6.5in]{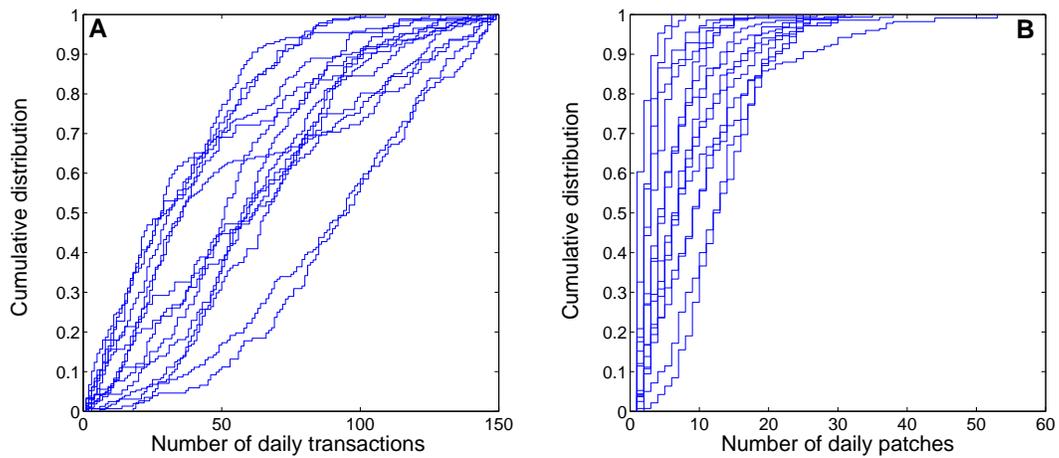}}
\caption{Individual trading activity. {\bf A} and {\bf B} show the cumulative distribution of the number of total transactions and number of patches visited each day by each trader. In our data, most of the traders typically ($>90\%$ of the time) made more than $10$ transactions and made more than $3$ switches per day.}
\label{fig2}
\end{figure*}

\begin{figure*}
\centerline{\includegraphics*[width=3.5in]{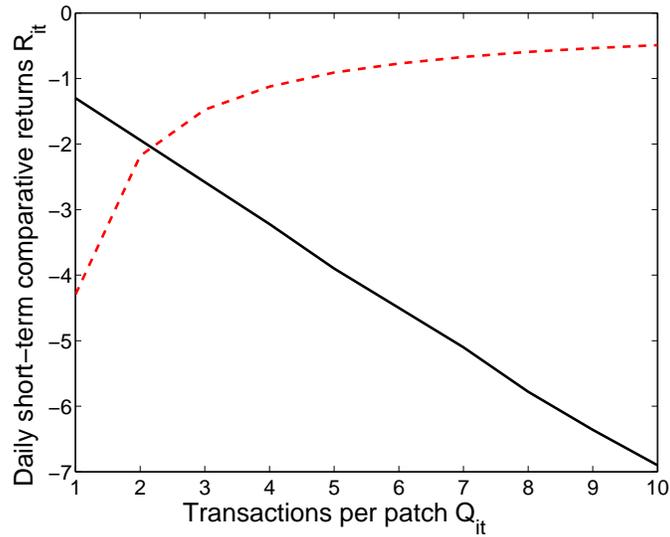}}
\caption{Importance of foraging choices over daily short-term returns. The figure shows an illustrative example of the relationship between daily or daily short-term comparative returns $\hat{R}_{it}$ and different exploration and exploitation choices as given by a constant number of transactions per patch $Q_{it}$. In the example, we considered a trader with $N_{it}=65$, $\beta_1=0.002$ and $\beta_2=0.02$. $\hat{R}_{it}=\sum_{j=1}^{j=65}\hat{z}_{ijt}$, where $\hat{z}_{ijt}$ are the predicted short-term comparative returns from the regression model (see text). When considering exploration only, the lower the value of $Q_{it}$ the higher the decline of $\hat{R}_{it}$ (dashed line); and the opposite behavior is observed when considering exploitation only (solid line). Importantly, this reveals that an optimal pattern for jointly maximizing traders' $\hat{R}_{it}$ exists, i.e. the intersection between the two curves.}
\label{fig3}
\end{figure*}

\begin{figure*}
\centerline{\includegraphics*[width=4in]{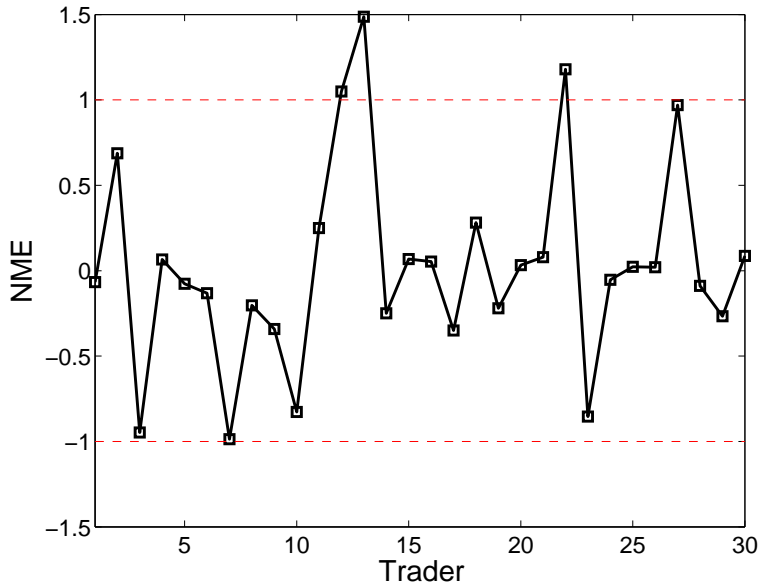}}
\caption{Optimal vs. real foraging choices. For each trader, the figure shows the normalized model errors (NME) between the optimal number of transactions per patch $Q^*_{ij}$ and the distribution of actual values of $Q_{ij}$. Here, the NME was computed as the difference between $Q^*_{ij}$ and the observed median value of $Q_{ij}$ divided by the difference between the observed median value and the observed value of $Q_{ij}$ at the $2.5\%$ or $97.5\%$ quantiles, depending on whether the optimal value is lower or larger than the observed median value. The NME makes no particular assumption about the distribution of observed values. NME values between $(-1,1)$ can be taken as cases where the optimal value is significantly similar to the observed values \cite{Pires}. We found only 3 cases with NME values greater than 1 (Fig. 4). Importantly, this number of cases falls within the number of rejections $[0,4]$ that one would expect with $95\%$ confidence from a Binomial model $B(30,0.05)$ \cite{Dean}. Thus, one cannot reject the hypothesis that this model is a good approximation to the observed exploration and exploitation choices of traders.
}
\label{fig4}
\end{figure*}

\begin{figure*}
\centerline{\includegraphics*[width=6.5in]{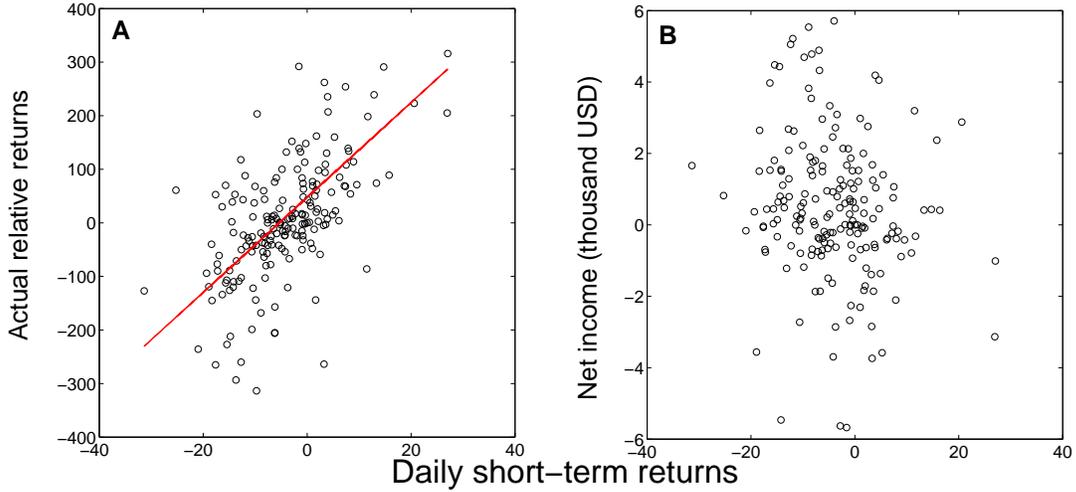}}
\caption{Association of daily short-term comparative returns with actual relative returns and net income intake. For illustrative purposes, {\bf A} shows the positive and significant association between daily short-term comparative returns and actual relative returns for one single trader and {\bf B} shows the association between daily short-term comparative returns and net income intake for the same trader. Correlation values for all traders are reported in Table 1. Daily short-term comparative returns are given by $R_{it}=\sum_{j=1}^{j=N_{it}}z_{ijt}$. Actual relative returns $A_{it}$ are calculated similar to the short-term comparative returns measure except that they do not take into account the standard deviation; and instead they multiply returns by the number of stocks sold or bought. Net income intake, $I_{it}$, is simply the amount of money made by traders; it is does not compare it with competitors. Each symbol corresponds to one trading day.
}
\label{fig5}
\end{figure*}

\end{document}